\documentclass{aa}

\usepackage{natbib}
\bibpunct{(}{)}{;}{a}{}{,} 

\usepackage{graphicx}

\usepackage{amssymb,amsmath}
\usepackage{delarray}
\usepackage{mathrsfs}
\usepackage{xcolor}
\usepackage{soul}
\usepackage[varg]{txfonts}
\usepackage{hyperref}
\hypersetup{
  colorlinks   = true, 
  urlcolor     = blue, 
  linkcolor    = blue, 
  citecolor    = blue  
}

\newcommand{\rr}{{\mathbf r}}

\begin{document} 

\title{Modeling the scattering polarization of the solar Ca~{\sc i} 4227~{\AA} line\\ with angle-dependent partial frequency redistribution}

\author{Gioele Janett\inst{1,2}
       \and
       Ernest Alsina Ballester\inst{1,3}
       \and
       Nuno Guerreiro\inst{1,2}
       \and
       Simone Riva\inst{1,2}
       \and
       Luca Belluzzi\inst{1,4,2}
       \and\\
       Tanaus\'u del Pino Alem\'an\inst{3}
       \and
       Javier Trujillo Bueno\inst{3,5,6}
       }

\institute{Istituto Ricerche Solari (IRSOL), Universit\`a della Svizzera italiana (USI), CH-6605 Locarno-Monti, Switzerland
          \and
          Euler Institute, Universit\`a della Svizzera italiana (USI), CH-6900 Lugano, Switzerland
          \and
          Instituto de Astrof\'isica de Canarias (IAC), E-38205 La Laguna, Tenerife, Spain
          \and
          Leibniz-Institut f\"ur Sonnenphysik (KIS), D-79104 Freiburg i.~Br., Germany
          \and
          Departamento de Astrofísica, Universidad de La Laguna, E-38206 La Laguna, Tenerife, Spain
          \and
          Consejo Superior de Investigaciones Científicas, Spain\\
          \email{gioele.janett@irsol.usi.ch}}
 
\abstract
{
The correct modeling of the scattering polarization signals observed in several strong 
resonance lines requires taking partial frequency redistribution (PRD) phenomena into account.
Modeling scattering polarization with PRD effects is very computationally demanding and the simplifying 
angle-averaged (AA) approximation is therefore commonly applied.
}
%
{
This work aims at assessing the impact and the range of validity of the AA approximation with respect to
the general angle-dependent
(AD) treatment of PRD effects in the modeling of scattering polarization
in strong resonance lines, with focus on the solar Ca~{\sc i} 4227~{\AA} line.
%
}
%
{
Spectral line polarization is modeled by solving the radiative transfer problem
for polarized radiation, under nonlocal thermodynamic equilibrium conditions, taking PRD effects into 
account, in static one-dimensional semi-empirical atmospheric models presenting arbitrary magnetic fields.
The problem is solved through a two-step approach.
In step 1, the problem is solved for intensity only, considering a multi-level atom. 
In step 2, the problem is solved including polarization,
considering a two-level atom with an unpolarized and infinitely sharp lower level,
and fixing the lower level population calculated at step 1.
}
%
{
The results for the Ca~{\sc i} 4227\,{\AA} line show 
a good agreement between the AA and AD calculations 
for the $Q/I$ and $U/I$ wings signals.
However, AA calculations reveal an artificial trough in the line-core peak of the linear polarization profiles,
whereas AD calculations show a sharper peak in agreement with observations.
}
%
{
An AD treatment of PRD effects is essential to correctly model the line-core peak of the 
scattering polarization signal of the Ca~{\sc i} 4227~{\AA} line.
By contrast, in the considered static case, the AA approximation seems to be 
suitable to model the wing scattering 
polarization lobes and their magnetic sensitivity through magneto-optical effects.
}

\keywords{Radiative transfer -- Scattering -- Polarization -- Sun: atmosphere -- Methods: numerical}

\titlerunning{Modeling the scattering polarization of the solar Ca~{\sc i} 4227~{\AA} line}
\authorrunning{Janett et al.}

\maketitle

\section{Introduction}\label{sec:sec1}
%
When observing quiet regions close to the edge of the solar disk (limb), 
several strong resonance lines, such as H~{\sc i} Ly-$\alpha$, Ca~{\sc ii} K, Ca~{\sc i} 4227, or Na~{\sc i} D$_2$, show significant 
scattering polarization signals, characterized by a relatively sharp peak in the line core and broader lobes in the wings \citep[e.g.,][]{gandorfer2000,gandorfer2002,kano2017}.
The observation and modeling of these polarization signals has been receiving increasing attention by the scientific 
community. 
In fact, their sensitivity to the presence of relatively weak magnetic fields
is presently one of the most promising tools to probe the magnetism
of the solar chromosphere and transition region \citep[e.g.,][]{trujillo_bueno2017}.
This magnetic sensitivity arises from the combined action of the Hanle, 
Zeeman, and magneto-optical effects \citep[e.g.,][]{alsinaballester2017}. 
The interest in these signals is clearly attested by the recent Chromospheric Ly-alpha Spectropolarimeter 
(CLASP) and Chromospheric Layer SpectroPolarimeter (CLASP2) sounding rocket experiments, which provided unprecedented 
spectropolarimetric observations of the H~{\sc i} Ly-alpha line
\citep{kano2017,trujillo_bueno2018} and of the Mg~{\sc ii} 
h and k lines \citep{ishikawa2021}, respectively.                            
In the visible and infrared ranges of the solar spectrum, scattering polarization signals of strong chromospheric lines can be
observed from the ground with instruments such as the Zurich Imaging Polarimeter \citep[ZIMPOL,][]{ramelli2010} or the 
Tenerife Infrared Polarimeter II \citep[TIP-II,][]{collados2007}, and will be among the targets of the new generation of large-aperture 
solar telescopes (e.g., DKIST and EST).

The scattering polarization signals of strong resonance lines are primarily produced by scattering
processes that are coherent in frequency in the atomic reference frame, and subject to Doppler redistribution in the observer's frame.
The core of these signals can be often suitably modeled under the limit of complete frequency redistribution 
\citep[CRD; e.g.,][]{sampoorna2010,stepan2015}. However, this approximation turns out to be completely inadequate to model their extended 
wing lobes, for which a general approach accounting for partial frequency redistribution (PRD) effects is of the 
utmost importance.
%
The numerical solution of the radiative transfer (RT) problem for polarized radiation under nonlocal thermodynamic equilibrium (NLTE) conditions, 
taking scattering polarization and PRD effects into account is a 
notoriously challenging problem from the computational point of view.
For this reason, the so-called angle-averaged (AA) simplifying assumption,
first introduced in the unpolarized case 
\citep[e.g.,][]{mihalas1978}, is still commonly applied 
to the modeling of 
scattering polarization signals
\citep[see, e.g.,][]{rees1982,belluzzi2014,alsinaballester2017}.
By smearing out geometrical aspects of the problem, the AA assumption can, however, introduce significant inaccuracies,
and it would be completely inadequate when aiming at modeling the aforementioned signals in realistic atmospheric models, 
which consider the detailed three-dimensional (3D) structure of the solar atmosphere. 

This highlights the need for methods that, exploiting the increased computational resources available today,
make it possible to model these signals by solving the NLTE RT problem for polarized radiation, taking PRD 
effects into account in the general angle-dependent (AD) case.
In fact, considerable efforts have already been put in this direction. \citet{dumont1977} pioneered the 
modeling of spectral line polarization accounting for AD PRD effects. 
Subsequently, \citet{nagendra2002} considered the AD PRD problem in the weak-field Hanle regime for the case of 
a two-level atom, using the theoretical 
approach of \citet{stenflo1994}, based on the Kramers-Heisenberg formula. 
\citet{sampoorna2008,sampoorna2017} generalized that work to the Hanle-Zeeman regime considering arbitrary magnetic fields.
\citet{sampoorna2011} and \citet{supriya2012} discussed the validity of the AA assumption in the presence of micro-turbulent magnetic fields, 
while \citet{supriya2013} extended the analysis to vertical magnetic fields. 
\citet{sampoorna2015apj} developed an AD PRD RT code capable of including vertical bulk velocities.
For computational simplicity, the AD PRD calculations presented in the aforementioned works 
were performed considering one-dimensional (1D) homogeneous atmospheric models. 
More recently, \citet{delpinoaleman2020} solved the AD PRD problem for a three-term atomic model in more realistic 1D atmospheric models,
applying the theoretical approach of \citet{casini2014,casini2017b}. 
Their code can take
into account the effect of AD PRD in the presence of magnetic fields of arbitrary strength and orientation, as well as vertical bulk velocities.
However, their AD PRD calculations were limited to
problems with axial symmetry,
because of the extreme computational cost.

%

This paper shows the results obtained
with a new code 
capable of solving the NLTE RT problem in static 1D semi-empirical models of the solar atmosphere,
taking scattering polarization and AD PRD effects into account.
Based on the theoretical approach developed by \citet{bommier1997a,bommier1997b},
the code allows considering two-level atomic models with an unpolarized and infinitely sharp lower level,
in the presence of arbitrary magnetic fields.
The present investigation is specifically focused on the Ca~{\sc i} 4227~{\AA} line,
which is an ideal test bench for new approaches to the modeling of scattering polarization
including PRD effects.
Indeed, the 
scattering polarization signal of this line clearly shows the impact of 
such effects \citep{faurobert_scholl1992} and can be suitably modeled by considering 
a simple two-level atom.
Furthermore, the Stokes profiles of this line are of high scientific
interest for magnetic field diagnostics, and
high-precision spectropolarimetric observations are available
\citep[e.g.,][]{bianda2011,supriya2014,alsinaballester2018,capozzi2020}.

The logical structure of this paper is as follows:
Section~\ref{sec:problem} outlines the formulation of the problem, exposing the 
relevant continuous equations and their discretization, and presenting a suitable iterative numerical approach 
to solve it. Next, Section~\ref{sec:results} focuses on the modeling of the Ca~{\sc i} 4227~{\AA} line, 
presenting a quantitative comparison between the results of 
AA and AD calculations. 
Finally, Section~\ref{sec:conclusions} summarizes the main results of this paper and discusses on their implications.

\section{Formulation of the problem and methods}\label{sec:problem}
In this section, 
we first present the continuous formulation of the considered NLTE 
RT problem for polarized radiation, including AD PRD effects. 
We then expose the discretization and the algebraic formulation of the 
problem and
outline the numerical iterative solution method that was applied.
In addition, 
we finally recall the simplifying AA approximation.
\subsection{Continuous formulation of the problem}\label{sec:continuous_problem}

The physical quantities entering the problem are, in general, functions of the
spatial point $\rr$, and of the frequency $\nu$ and propagation direction 
$\vec{\Omega}$ of the radiation beam under consideration.
A complete description of the polarization state of the radiation field is 
provided by the four Stokes parameters
$$I_i(\rr,{\bf\Omega},\nu),$$
with $i=1,\ldots,4$, standing for Stokes $I$, $Q$, $U$, and $V$, respectively.
The transfer of partially polarized light along direction $\vec{\Omega}$ 
at frequency $\nu$ is described by the system of coupled first-order
inhomogeneous ordinary differential equations given by
\begin{equation}
  \frac{\rm d}{{\rm d} s}I_i(\rr,\vec{\Omega},\nu)  = -\sum_{j=1}^4 K_{ij}(\rr,{\bf\Omega},\nu)I_j(\rr,{\bf\Omega},\nu) + \varepsilon_i(\rr,{\bf\Omega},\nu),
\label{eq:RTE}
\end{equation}
where $s$ is the spatial coordinate measured along the direction $\vec{\Omega}$, 
$K_{ij}$ are the elements of the so-called propagation matrix
\begin{equation*}
  K = \begin{pmatrix}
      \eta_1 &  \eta_2 &  \eta_3 & \eta_4  \\
      \eta_2 &  \eta_1 &  \rho_4 & -\rho_3 \\
      \eta_3 & -\rho_4 &  \eta_1 & \rho_2  \\
      \eta_4 &  \rho_3 & -\rho_2 & \eta_1 
      \end{pmatrix},
\label{matrix_K}
\end{equation*}
where $\eta_i$ and $\rho_i$ are the dichroism coefficients and the anomalous dispersion coefficients, respectively,
while $\varepsilon_i$ are the emission coefficients in the four Stokes parameters.
In general, $K_{ij}$ and $\varepsilon_i$
contain contributions due to both line and continuum processes
and can be expressed as
\begin{align}
    \label{eq:prop_mat_lc}
    K_{ij}(\vec{r},\vec{\Omega},\nu) & = K_{ij}^{\ell}(\vec{r},\vec{\Omega},\nu) + 
    K_{ij}^c(\vec{r},\nu), \\
    \label{eq:emis_vec_lc}
    \varepsilon_i(\vec{r},\vec{\Omega},\nu) & = \varepsilon_i^{\ell}(\vec{r},\vec{\Omega},\nu) 
    +\varepsilon_i^{c}(\vec{r},\vec{\Omega},\nu),
\end{align}
with the superscripts $\ell$ and $c$
standing for line and continuum, respectively.

In this work, we consider a two-level atomic model
with an unpolarized and
infinitely sharp lower level in the presence of magnetic fields of arbitrary
intensity and orientation.
For such an atomic model, an analytical solution
of the statistical equilibrium equations is available
and the line contribution to the emission coefficient can be 
thus expressed through the redistribution matrix formalism. 
Stimulated emission is negligible
around the Ca~{\sc i} 4227~{\AA} frequency
in the solar atmosphere
and is consequently neglected.
Moreover, bulk velocity fields are not considered;
their impact on this problem
will be the subject of an upcoming investigation.
\subsubsection{Line contributions}
The line contributions to $\eta_i$ and $\rho_i$ for the considered atomic model are given by 
\citep[see][]{alsinaballester2017}
\begin{align}
\eta_i^\ell(\rr,\vec{\Omega},\nu) & = k_L(\rr) \sum_{K=0}^2 \hat{\mathcal{T}}^K_{0,i}(\rr,\vec{\Omega})
	\Phi^{0K}_0(\rr,\nu),\\
\rho_i^\ell(\rr,\vec{\Omega},\nu) & = k_L(\rr) \sum_{K=0}^2 \hat{\mathcal{T}}^K_{0,i}(\rr,\vec{\Omega})
	\Psi^{0K}_0(\rr,\nu), 
\end{align}
where $k_L$ is the frequency-integrated absorption coefficient, which depends on the population of the 
lower level, $\hat{\mathcal{T}}^K_{Q,i}$ is the polarization tensor 
\citep[see][Chapter 5]{landi_deglinnocenti+landolfi2004} evaluated in the magnetic reference 
system\footnote{It is the reference system with the $z$-axis directed along the local magnetic field vector, which
depends on the spatial point $\rr$.
This explains the dependence of the polarization tensor $\hat{\mathcal{T}}^K_{Q,i}$ on $\rr$.},
while $\Phi^{KK'}_Q$ and $\Psi^{KK'}_Q$ are the generalized profiles defined in 
\citet[][Appendix 13]{landi_deglinnocenti+landolfi2004}.

Within the redistribution matrix formalism, one can distinguish between thermal and scattering
contributions to the line emission coefficients, namely,
\begin{equation*}
\varepsilon_i^\ell(\rr,{\bf\Omega},\nu)=\varepsilon_i^{\ell,th}(\rr,{\bf\Omega},\nu)+\varepsilon_i^{\ell,sc}(\rr,{\bf\Omega},\nu).
\end{equation*}
Assuming that inelastic collisions are isotropic,
the thermal contribution to the line emission coefficients
for the considered atomic system 
reads \citep[see][]{alsinaballester2017}
\begin{equation}
\begin{aligned}
\varepsilon_i^{\ell,th}(\rr,\vec{\Omega},\nu) &= k_L(\rr) \epsilon(\rr)
	W_T(\rr)\sum_{K=0}^2 \hat{\mathcal{T}}^K_{0,i}(\vec{\Omega})
	\Phi^{0K}_0(\rr,\nu)\\
	&=\epsilon(\rr)W_T(\rr)\eta_i^\ell(\rr,\vec{\Omega},\nu),
\end{aligned}
\end{equation}
where $\epsilon$ is the thermalization parameter, and $W_T$ is the Planck 
function in the Wien limit at the line-center frequency $\nu_0$.
The scattering contribution to the line emission coefficients reads
{\small
\begin{equation}\label{eq:epsilon_sc}
	\varepsilon_i^{\ell,sc}(\rr,{\bf\Omega},\nu)\!=\!
	k_L(\rr)\!\int \!{\rm d} \nu'\! \oint\! \frac{{\rm d} {\bf\Omega}'}{4 \pi}\sum_{j=1}^4
	R_{ij}(\rr,{\bf\Omega},{\bf\Omega}',\nu, \nu')
	I_j(\nu',{\bf\Omega}',\rr),
\end{equation}}\noindent 
where $R_{ij}$ are the elements of the redistribution matrix and $I_j$ are the Stokes parameters of the radiation 
that pumps the atomic system.
Hereafter, the primed and unprimed quantities refer to the incoming and outgoing radiation, respectively.
The redistribution matrix derived by \citet{bommier1997b} is given by the sum of two terms
\begin{equation*}
R_{ij}(\rr,{\bf\Omega},{\bf\Omega}',\nu, \nu')=
R^{\scriptscriptstyle \rm II}_{ij}(\rr,{\bf\Omega},{\bf\Omega}',\nu, \nu')+
R^{\scriptscriptstyle \rm III}_{ij}(\rr,{\bf\Omega},{\bf\Omega}',\nu, \nu'),
\end{equation*}
where $R^{\scriptscriptstyle \rm II}$ and $R^{\scriptscriptstyle \rm III}$
describe scattering processes 
that are coherent and completely uncoherent, respectively, in the atomic reference frame. 
The transformation of 
$R^{\scriptscriptstyle \rm II}$  
of \citet{bommier1997b} 
from the atomic to the observer's frame 
(accounting for the Doppler effect due to thermal motions)
is described in detail in \citet{alsinaballester2017}.
The assumption of CRD in the observer's frame is instead made for $R^{\scriptscriptstyle \rm III}$ 
\citep[e.g.,][]{bommier1997b,alsinaballester2017}.
This approximation appears to be suitable for modeling strong chromospheric lines, whose core forms at atmospheric heights 
where the rate of elastic collisions is relatively low, and the contribution of $R^{\scriptstyle \rm III}$ is therefore 
much smaller than that of $R^{\scriptstyle \rm II}$. 
A quantitative analysis of the impact of this approximation on our modeling goes beyond the scope of this work and is left 
for a future investigation. 
The reader is referred to \citet{bommier1997b} and \citet{sampoorna2017} for more details on the suitability of this assumption.
The ensuing redistribution matrices can be written as
\begin{align*}
	R^{\scriptscriptstyle \rm II}_{ij}(\rr,{\bf\Omega},{\bf\Omega}',\nu, \nu') &=
	\sum_{KK'Q} \mathcal{R}^{{\scriptscriptstyle \rm II},KK'}_{Q}(\rr,\nu, \nu', \Theta)P^{KK'}_{Q,ij}({\bf\Omega},{\bf\Omega}'),\\
	R^{\scriptscriptstyle \rm III}_{ij}(\rr,{\bf\Omega},{\bf\Omega}',\nu, \nu') &=
	\sum_{KK'Q} \mathcal{R}^{{\scriptscriptstyle \rm III},KK'}_{Q}(\rr,\nu, \nu')P^{KK'}_{Q,ij}({\bf\Omega},{\bf\Omega}'),
\end{align*}
where $\Theta$ represents the scattering angle, that is the angle between the incoming ${\bf\Omega}'$
and outgoing ${\bf\Omega}$ directions.
The redistribution functions $\mathcal{R}^{{\scriptscriptstyle \rm II},KK'}_{Q}$
and $\mathcal{R}^{{\scriptscriptstyle \rm III},KK'}_{Q}$ and the elements of the scattering 
phase matrix $P^{KK'}_{Q,ij}$ are also described in \citet{alsinaballester2017}.

\subsubsection{Continuum contributions}\label{sec:continuum}
In the visible part of the solar spectrum, continuum processes only contribute significantly
to the diagonal element $\eta_I$ of the propagation matrix, namely,
\begin{align*}
\eta_i^c(\rr,\nu) & = k_c(\rr,\nu)\delta_{i1},\\
\rho_i^c & = 0,
\end{align*}
where $\delta_{ij}$ is the Kronecker delta.
The continuum contribution to the emission coefficients can be divided into a (unpolarized and isotropic) thermal term 
and a scattering term
\begin{equation}
    \varepsilon_i^c(\rr,{\bf\Omega},\nu)=\varepsilon_I^{c,th}(\rr,\nu) \delta_{i1} + 
    \varepsilon_i^{c,sc}(\rr,{\bf\Omega},\nu).
\end{equation}
The scattering contribution to the continuum emission coefficients reads \citep[see][]{alsinaballester2017}
{\small
\begin{equation}\label{eq:epsilon_sc_cont}
	\varepsilon_i^{c,sc}(\rr,{\bf\Omega},\nu)\!=\!
	\sigma(\rr,\nu)\!\sum_{KQ}(-1)^Q\mathcal{T}_{Q,i}^K(\mathbf{\Omega})
	\oint\! \frac{{\rm d} {\bf\Omega}'}{4 \pi}
	\sum_{j=1}^4\mathcal{T}_{-Q,j}^K(\mathbf{\Omega})
	I_j(\mathbf{r},\mathbf{\Omega},\nu),
\end{equation}}\noindent
where $\sigma$ is the continuum opacity for scattering and
$\mathcal{T}^K_{Q,i}$ is the polarization tensor 
\citep[see][Chapter 5]{landi_deglinnocenti+landolfi2004} evaluated in the reference 
system of the problem\footnote{It is the reference system with the $z$-axis directed along the vertical of the atmospheric model.}.

In summary, in order to determine
the Stokes parameters $I_i$ of the radiation emerging from the considered atmospheric model, one needs to solve the transfer equation~\eqref{eq:RTE}. This requires the knowledge of the emission vector $\varepsilon_i$ which, in turn, depends on the Stokes parameters 
$I_i$ through~\eqref{eq:epsilon_sc} and \eqref{eq:epsilon_sc_cont}. 
The problem thus consists in finding a self-consistent solution for $I_i$ and $\varepsilon_i$.
%

\subsection{Discretization and algebraic formulation}\label{sec:algebraic_formulation}
After a suitable discretization of the continuous problem \citep[see, e.g.,][]{janett2021a},
the solution of the transfer equation~\eqref{eq:RTE} and the calculation of the emission coefficients 
\eqref{eq:emis_vec_lc} can be expressed in the compact matrix form
\begin{align}
\mathbf{I}&=\Lambda\pmb{\varepsilon}+\mathbf{t},&&\;\;\text{ with }
\Lambda\in\mathbb R^{4  N_s N_\Omega N_\nu \times4 N_s N_\Omega N_\nu},
\label{matricial_form_1}\\
\pmb{\varepsilon}&=\Sigma\mathbf{I}+\mathbf{c},&&\;\;\text{ with }
\Sigma\in\mathbb R^{4  N_s N_\Omega N_\nu \times4 N_s N_\Omega N_\nu},
\label{matricial_form_2}
\end{align}
respectively, with
the vectors $\mathbf{I}$, $\pmb{\varepsilon}$, $\mathbf{t}$, $\mathbf{c}\in\mathbb R^{4 N_s  N_\Omega N_\nu}$.
$\mathbf{I}$ collects the discretized Stokes parameters and $\pmb{\varepsilon}$ 
the discretized emission coefficients. 
Moreover, $\Lambda\pmb{\varepsilon}$ describe the transfer of the radiation generated inside the
atmospheric model, $\mathbf{t}$ represents the radiation transmitted from the boundaries,
while $\Sigma\mathbf{I}$ and $\mathbf{c}$ represent the scattering and thermal contributions
to the emission coefficients, 
respectively.

By choosing $\mathbf{I}$ as the problem's unknown,
Equations~\eqref{matricial_form_1}--\eqref{matricial_form_2}
can then be combined in a single discrete problem, namely,
\begin{equation}\label{linear_system}
 (Id-\Lambda\Sigma)\mathbf{I}=\Lambda\mathbf{c}+\mathbf{t},
\end{equation}
where $Id$ is the identity matrix of size $4  N_s N_\Omega N_\nu$.
The discrete problem~\eqref{linear_system} is generally nonlinear,
because the coefficients of $\Lambda$ and $\mathbf{t}$ depend
in a nonlinear way on the 
absorption coefficient $\eta_I$,
which depends on the lower level atomic population,
which in turn is affected by the unknown $\mathbf I$.
However, if the lower level atomic population is fixed a priori,
the absorption coefficient $\eta_I$ becomes a constant of the problem.
In this case, $\Lambda$ and $\mathbf{t}$
become independent from the unknown $\mathbf I$
and the discrete problem~\eqref{linear_system} becomes linear.
\subsection{Numerical iterative solution}\label{sec:numerical_approach}
%
The numerical solution of the full NLTE RT problem is divided into two main steps:
\begin{enumerate}
\item The nonlinear NLTE RT problem is first solved
considering a multi-level atomic model, 
taking PRD effects into account, but neglecting
polarization and the effect of magnetic fields.
An accurate estimate of the populations of the atomic levels is thus provided.

\item A two-level atomic model with an unpolarized and infinitely sharp lower level is considered. 
The population of the lower level obtained in the first step is used and kept fixed.
The resulting linear problem~\eqref{linear_system} is solved in the general AD case. 
Although this problem can in principle be solved in a single step with a direct method,
this task is beyond current computational capabilities.
A stationary or Krylov iterative method
can therefore be applied \citep[see, e.g.,][]{janett2021a,benedusi2021}, using the unpolarized 
radiation field obtained in the first step as an initial guess.
\end{enumerate}
Although it is not fully consistent, this two-step approach permits to obtain
an accurate estimate of the lower level population from a multilevel atomic
model, while retaining the simplicity of a two-level atom when accounting for
polarization.
Moreover, by fixing the lower level population, the line absorption coefficient
becomes a constant, and the problem at
step 2 becomes linear (see Section~\ref{sec:algebraic_formulation}).
It is worth observing that the population of the upper level is
not kept fixed at step 2. This allows taking into account the impact of atomic
polarization (neglected at step 1) on the ratio between the population of the
upper and lower level.
In this respect, it must be observed that the degree of atomic polarization
is in general quite low, and therefore this impact is in most cases very small.
The validity of the approach is corroborated by previous works 
\citep[e.g.,][]{belluzzi2012,belluzzi2012b,alsinaballester2016,alsinaballester2018}.
\subsection{Angle-averaged approximation}\label{sec:AA}
In the general AD approach, one considers the inherent coupling between incoming and outgoing 
directions and frequencies, which manifests in presence of the scattering angle $\Theta$
in the redistribution function $\mathcal{R}^{{\scriptscriptstyle \rm II},KK'}_{Q}$.
Accurate AD calculations are computationally very expensive, and the AA approximation 
\citep[e.g.,][]{mihalas1978,rees1982} was thus proposed to simplify the numerical 
problem. 
According to the formulation outlined in Sect.~\ref{sec:continuous_problem},
this approximation consists in averaging 
the expression of $\mathcal{R}^{{\scriptscriptstyle \rm II},KK'}_{Q}$ 
over the scattering angle $\Theta$, namely,
\begin{equation*}
	\mathcal{\bar R}^{{\scriptscriptstyle \rm II},KK'}_{Q}(\rr,\nu, \nu') = \frac{1}{2}\int_0^\pi {\rm d} {\Theta}\sin(\Theta)
	\mathcal{R}^{{\scriptscriptstyle \rm II},KK'}_{Q}(\rr,\Theta,\nu,\nu').
\end{equation*}
The AA redistribution function $\mathcal{\bar R}^{{\scriptscriptstyle \rm II},KK'}_{Q}$
is thus independent of the scattering angle, partially decoupling (in the absence of bulk velocities) frequencies
and directions in the calculation of the line emissivity (see Equation~\eqref{eq:epsilon_sc}).
This allows for a drastic reduction of the computational
cost \citep[e.g.,][]{alsinaballester2017}.

This approximation proved suitable to model
the intensity profiles \citep[e.g.,][]{leenaarts2012,sukhorukov2017}.
In the modeling of polarization signals,
the range of validity of this approximation was analyzed by 
\citet{faurobert1987,faurobert1988} for the nonmagnetic case, by \citet{sampoorna2011} for weak magnetic fields, and by \citet{sampoorna2017} for arbitrary magnetic fields.
The overall conclusion is that the AA approximation
does not allow a satisfactory modeling of the 
linear polarization signals in the core of spectral lines.
These analyses have been, however, limited to academic scenarios, with hypothetical lines 
and homogeneous atmospheric models.
A rigorous comparison between the AA and the AD approaches for realistic atomic and atmospheric models
is still lacking and is the main scope of the next section.
%
\section{Numerical results}\label{sec:results}
This section presents the numerical modeling of the linear scattering 
polarization profiles of the Ca~{\sc i} line at 4227~{\AA}, as an illustrative application of the 
problem's formulation, discretization, and solution method described in Section~\ref{sec:problem}.
The calculations are carried out in 1D semi-empirical models of the 
solar atmosphere, in the presence of arbitrary magnetic fields, and taking AD PRD effects into account.
\subsection{The Ca~{\sc i} 4227~{\AA} spectral line}
%

The Ca~{\sc i} 4227\,{\AA} line is produced by the transition between the ground level of neutral calcium 
($4s^2 \, ^1{\rm S}_0$) and the excited level $4s4p \, ^1{\rm P}^{\rm o}_1$.
Since both levels pertain to singlets and the upper one is not radiatively connected to other levels with 
lower energy, the scattering polarization signal of this line
can be suitably modeled by considering a two-level atomic model.

The intensity spectrum of this line shows extended damping wings and a nearly saturated line core. 
For a disk center line of sight (LOS) and considering the
semi-empirical model~C of \citet{fontenla1993}, the line core forms at a height of about 
900\,km above the $\tau=1$ surface, with $\tau$ the continuum optical depth
at 5000\,{\AA} along the vertical. 
When observed in quiet regions close to the limb, this line exhibits the largest scattering polarization 
signal in the visible range of the so-called Second Solar Spectrum 
\citep[e.g.,][]{stenflo1980,stenflo1982,gandorfer2002}.
This signal presents a clear triple-peak structure, typical of strong resonance lines, with a 
sharp peak in the line core and broad lobes in the wings. 
The wing lobes form at photospheric heights (at around 180~km for an observation at $\mu=\cos{\theta}=0.1$, 
with $\theta$ the heliocentric angle), and can only 
be suitably modeled by taking PRD effects into account. 
The scattering polarization signal of this line is of particular interest because of its sensitivity 
to the magnetic fields present across a wide range of atmospheric heights. 
Indeed, the line-core peak is sensitive to low-chromospheric magnetic fields through the Hanle effect 
\citep{stenflo1982,faurobert_scholl1992}, while the wing lobes are sensitive to photospheric fields via 
magneto-optical effects \citep{alsinaballester2018}. 
Furthermore, the well-known Zeeman effect can be exploited as well in this line.

A considerable effort was indeed directed to the observation and modeling of
the scattering polarization signal of the Ca~{\sc i} 4227~{\AA} line.
Worth of mention are the observations of \citet{bianda2003}, which showed unexpected 
$U/I$ wing signals, as well as clear spatial variations of the amplitude of
both the $Q/I$ and $U/I$ wing lobes.
These observations could be explained in terms of magneto-optical effects by 
\citet{alsinaballester2018}.
This theoretical interpretation has gained further support from the 
observational work by \citet{capozzi2020}.
In the core of this line, scattering polarization signals can also be observed at the 
disk center,
the interpretation of which requires
3D radiative transfer modeling \citep[see][]{jaume_bestard2021}.
Observations of such disk-center signals were presented by \citet{bianda2011}, and subsequently analyzed by
\citet{anusha2011} to infer the magnetic fields present in the low chromosphere
under the assumption that the observed signals are due to the Hanle effect
in forward-scattering geometry \citep[see][]{trujillo_bueno2001}.
More recently, \citet{carlin2017} theoretically studied the temporal evolution of
these polarization signals using
the CRD and 1.5D approximations, emphasizing their sensitivity to gradients in the vertical component of the bulk velocities, while
\citet{jaume_bestard2021} investigated the linear and circular polarization signals of
the Ca~{\sc i} 4227~{\AA} line in full 3D, pointing out the important symmetry 
breaking produced by the spatial gradients of the horizontal component
of the macroscopic velocities.
Moreover, \citet{supriya2014} modeled the center-to-limb variation of the $Q/I$ 
profile of this line, and compared it to observations recorded with the ZIMPOL-III instrument \citep[e.g.,][]{ramelli2010}.

\subsection{Numerical setting and methods}
The Stokes profiles of the Ca~{\sc i} 4227~{\AA} line presented
in the following sections were 
calculated in the 1D semi-empirical models of \citet[][hereafter FAL models]{fontenla1993}, whose spatial grids 
contain between 60 and 70 nodes.
A Cartesian reference system with the $z$-axis (quantization axis
for total angular momentum) directed along the vertical is considered. 
A given propagation direction $\vec{\Omega}$ is specified through its inclination $\theta \in [0,\pi]$ 
(i.e., $\mu=\cos{\theta} \in [-1,1]$) with respect to the vertical, and azimuth $\chi \in [0,2\pi)$ measured 
counter-clockwise from the $x$-axis.
The angular grid
is given by the tensor product of 8 equally spaced nodes in the azimuthal interval $[0,2\pi]$,
and 6 Gauss-Legendre quadrature nodes in each inclination subinterval $[-1,0]$ 
and $[0,1]$, for a total of 96 nodes.
The wavelength interval $[4071 \text{\,\AA},4261 \text{\,\AA}]$ is discretized 
onto a spectral grid with 155 nodes, 
equally spaced in the line core and logarithmically distributed in the wings.
In all calculations, the reference direction for positive 
Stokes $Q$ is taken 
parallel to the limb.

The whole problem is solved through the two-step approach presented in Section~\ref{sec:numerical_approach}.
The first step is carried out by means of the RH code \citep{uitenbroek2001},
using the parabolic short-characteristic formal solver.
In step 1, an atomic model for calcium with 25 levels (19 Ca~{\sc i} levels, 
5 Ca~{\sc ii} levels, and the ground level of Ca~{\sc iii}) is considered. 
This model accounts for 21 line transitions and 24 continuum transitions.
All spectral lines are computed in the CRD limit,
except for the Ca~{\sc i} 4227~{\AA} line and the Ca~{\sc ii} H \& K lines, which receive a PRD treatment.
The RH code is also used to calculate 
the continuum total opacity $k_c$, scattering opacity $\sigma$, 
and thermal emissivity $\varepsilon_I^{c,th}$, as well as the collisional rates. 
The rates for inelastic collisions with electrons, inducing transitions between the upper and lower levels 
of the Ca~{\sc{i}} 4227~{\AA} line, 
are computed according to the formula given in \cite{Seaton62}. 
The rates of elastic collisions
are calculated including the contributions from the 
Van der Waals interaction with neutral hydrogen and helium atoms, computed following \cite{Unsold55}, 
and from the quadratic Stark effect, due to electrons and singly charged ions \citep{Traving60, Gray92}.  

Step 2 is carried out by means of a Matlab code that solves the linear problem~\eqref{linear_system} 
in the general AD case 
with a Richardson iterative method \citep[see][]{janett2021a},
using the unpolarized radiation field calculated at step 1 
as initial guess. In step 2, a two-level atomic model is considered. 
The population of the lower level is fixed to the value calculated in the first step and 
left unchanged. The DELO-parabolic formal solver \citep{janett2017a,janett2018b} is used
to solve the transfer equation~\eqref{eq:RTE}.

Concerning the choice of the formal solver, it must be recalled that
high-order schemes generally outperform low-order methods if the
atmospheric model guarantees sufficient smoothness in the solution. 
This turns out to be the case for the FAL models, and
for this reason, the third-order accurate DELO-parabolic method of \citet{janett2017a}
was used to perform the calculations presented in this paper. 
Even though \citet{janett2018b} did not recommend multistep schemes for practical applications,
the smoothness of the FAL atmospheric models allows the safe use of the 
DELO-parabolic method, guaranteeing its proximity to $L$-stability \citep[see][]{janett2018a}.

In order to facilitate the convergence in step 2, 
an intermediate step can be added.
In this additional step, the linear problem~\eqref{linear_system}
is solved with a stationary iterative method
under the AA approximation \citep[see][]{alsinaballester2017}.
This calculation provides temporary solution for the polarized radiation field, 
which acts as a better initial guess for the step 2.
At the same time, the Stokes profiles resulting from the intermediate step also provides a comparison suitable to 
identify the impact and the range of validity of the AA approximation
with respect to the general AD treatment of PRD effects.
These AA calculations
are carried out using a modified version of the Fortran code developed by \citet{alsinaballester2017}, which 
solves the linear problem (12) taking PRD effects into account under the AA approximation.
Unlike the original version of the code, this modified version uses a damped Jacobi iterative method\footnote{The idea of damping consists in
multiplying the correction in the Jacobi iteration by a damping parameter
(or relaxation factor) $\omega$, which typically satisfies $0<\omega<1$ 
\citep[see, e.g.,][]{hackbusch1994,janett2021a}.} and the DELO-parabolic formal solver. 
The damped Jacobi method is used because the Jacobi iterative method does not 
guarantee convergence when used in combination with the DELO-parabolic formal solver 
\citep[see][]{janett2021a}. A detailed analysis on the convergence properties of various 
iterative methods applied to transfer problems of polarized radiation can be found in
\citet{janett2021a}.
%

\subsection{Nonmagnetic case}
The impact of the AD treatment on the modeling of the emergent Stokes profiles 
of the Ca~{\sc i} 4227~{\AA} line 
is first analyzed in the absence of magnetic fields, 
considering the four different FAL models.
These atmospheric models represent a faint inter-network region of the quiet Sun (FAL-A),
an average region of the quiet Sun (FAL-C), a bright network region of the quiet Sun (FAL-F),
and a typical plage area (FAL-P).

\begin{figure}
\includegraphics[width=0.49\textwidth]{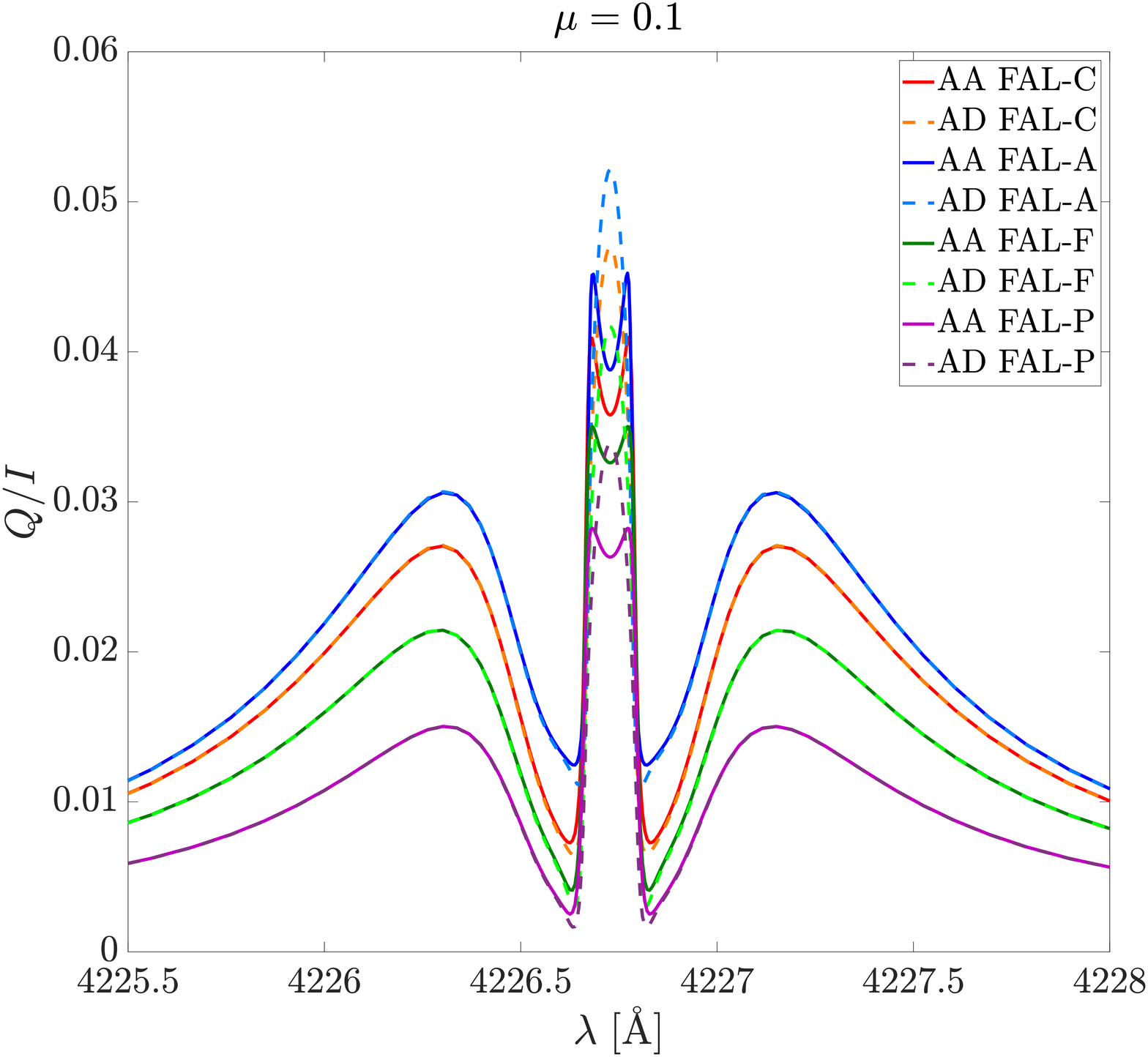}
\caption{Emergent $Q/I$ profiles calculated in different atmospheric models 
for a LOS with $\mu = 0.1$, taking into account PRD effects both under the AA approximation 
(solid lines) and in the general AD case (dashed lines). 
No magnetic field is considered. The considered atmospheric models are: FAL-C (red/orange lines),
FAL-A (blue/light blue lines), FAL-F (green/pale green lines), and FAL-P (purple/violet lines).
The reference direction for positive $Q$ is taken parallel to the limb.}
\label{fig:atmos_models}
\end{figure}

Figure~\ref{fig:atmos_models} shows the emergent $Q/I$ profiles for the 
aforementioned 
atmospheric models and a 
LOS with $\mu=0.1$.
In the absence of magnetic fields, the $U/I$  
profiles vanish and are consequently not shown.
An interesting result is the following:
the small trough in the line-core peak of the $Q/I$ profile,
obtained 
when the AA approximation is considered, completely disappears 
when a general AD calculation is carried out.
The trough found in the AA case can be thus attributed to
the averaging over the scattering angle explained in Section~\ref{sec:AA}.
This clearly highlights the scientific value of AD calculations, as 
the sharper $Q/I$ peak attained in this case is in much 
better agreement with the observations \citep[e.g.,][]{gandorfer2002}. 
By contrast, a good agreement between AA and AD calculations
is found in the wings of the $Q/I$ profile.
%

\begin{figure}
\includegraphics[width=0.49\textwidth]{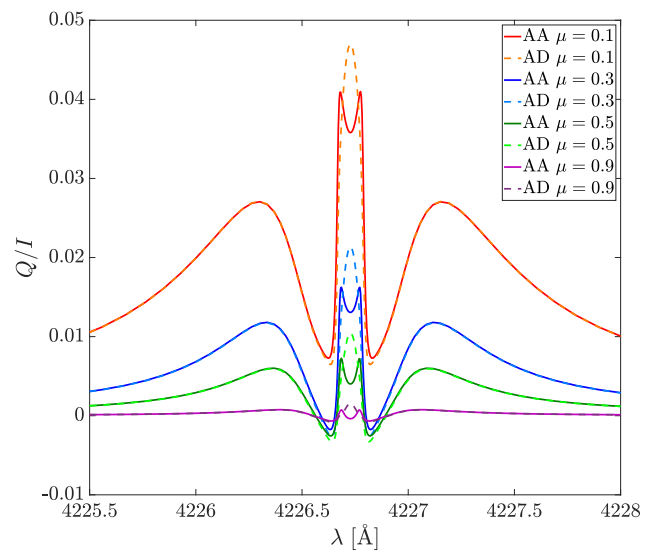}
\caption{Emergent $Q/I$ profiles calculated in the FAL-C atmospheric model
for different LOS, taking into account PRD effects both under the AA approximation 
(solid lines) and in the general AD case (dashed lines). No magnetic field is considered.
The considered LOS are: $\mu = 0.1$ (red/orange lines), $\mu = 0.3$ (blue/light blue lines), 
$\mu = 0.5$ (green/pale green lines), and $\mu = 0.9$ (purple/violet lines).
The reference direction for positive $Q$ is taken parallel to the limb.}
  \label{fig:center_to_limb}
\end{figure}

AA and AD calculations are now compared considering the FAL-C atmospheric model 
in the absence of magnetic fields for different LOS.
The results shown in Figure~\ref{fig:center_to_limb} are consistent with
those of Figure~\ref{fig:atmos_models}.
For all the considered LOS, AA calculations yield a trough in the line-core 
peak of the $Q/I$ profile, whereas the AD counterparts show a peak.
Additionally, AA and AD calculations 
substantially agree in the wings of the $Q/I$ profile.
As expected, the amplitude of the $Q/I$ profile, both in the core and in the 
wings, decreases when moving towards the disk center (i.e., for increasing $\mu$ values).

\subsection{Magnetic case}
%
%
The linear scattering polarization signal of the Ca~{\sc i} 4227\,{\AA} line is 
primarily sensitive to the presence of magnetic fields through the combined action 
of the Hanle and magneto-optical effects.
The Hanle effect operates in the line core, causing a reduction of the degree of 
linear polarization and a rotation of the plane of linear polarization. 
The Hanle critical field for this line is of approximately 25\,G. 
Magneto-optical effects mainly operate in the line wings, producing a rotation of the 
plane of linear polarization, and a reduction of the linear polarization degree.
The impact of this mechanism becomes appreciable in the presence of longitudinal magnetic fields
with strengths similar to those required for the Hanle effect to have a substantial impact
\citep[more details in][]{alsinaballester2018,alsinaballester2019}.
The polarization of the Ca~{\sc i} 4227~{\AA} line is also sensitive to the well-known 
Zeeman effect, which produces measurable circular polarization signals in the presence 
of magnetic fields with strengths larger than 10\,G in the upper solar photosphere.

\begin{figure*}
 \includegraphics[width=0.49\textwidth]{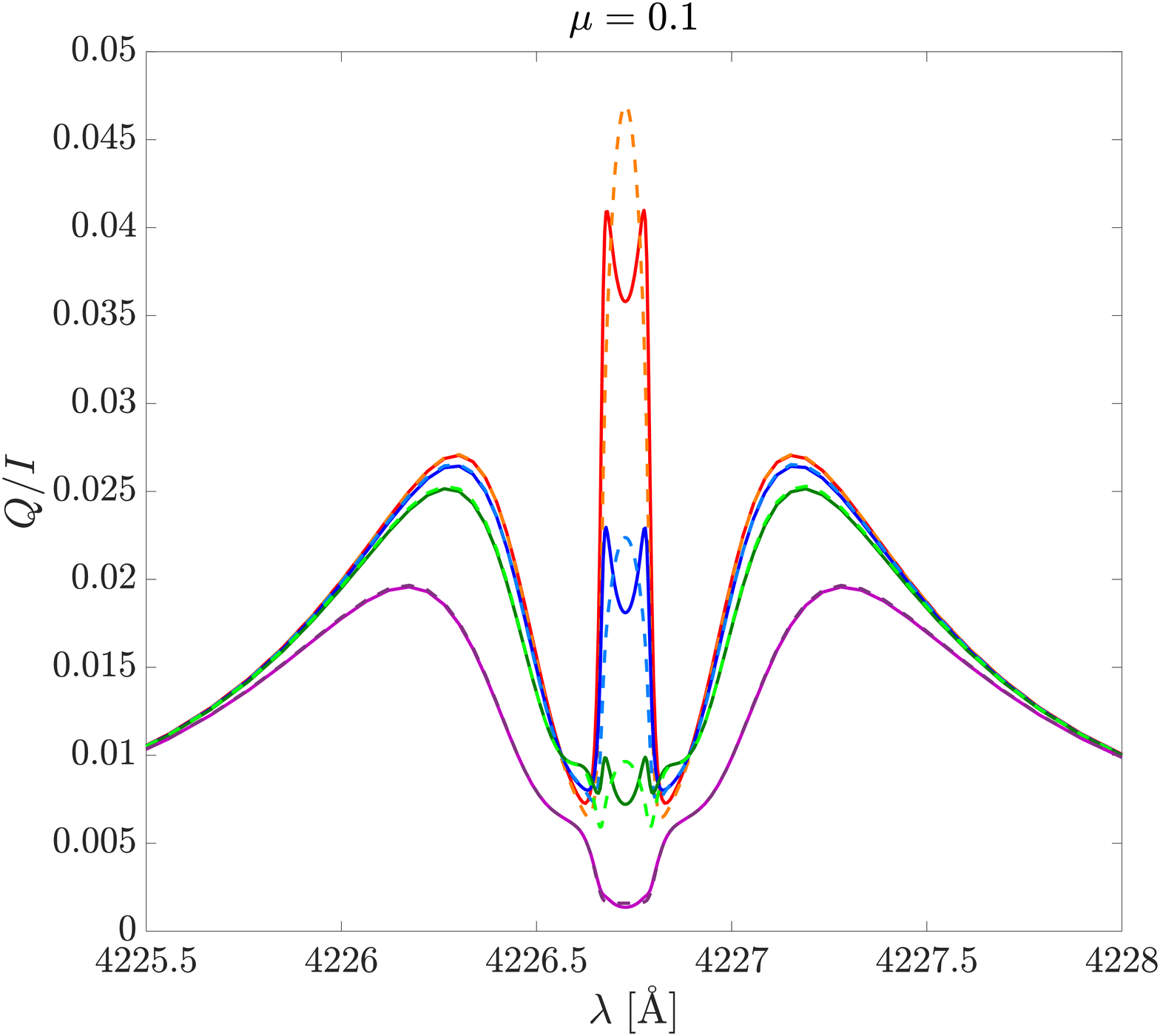}
 \includegraphics[width=0.49\textwidth]{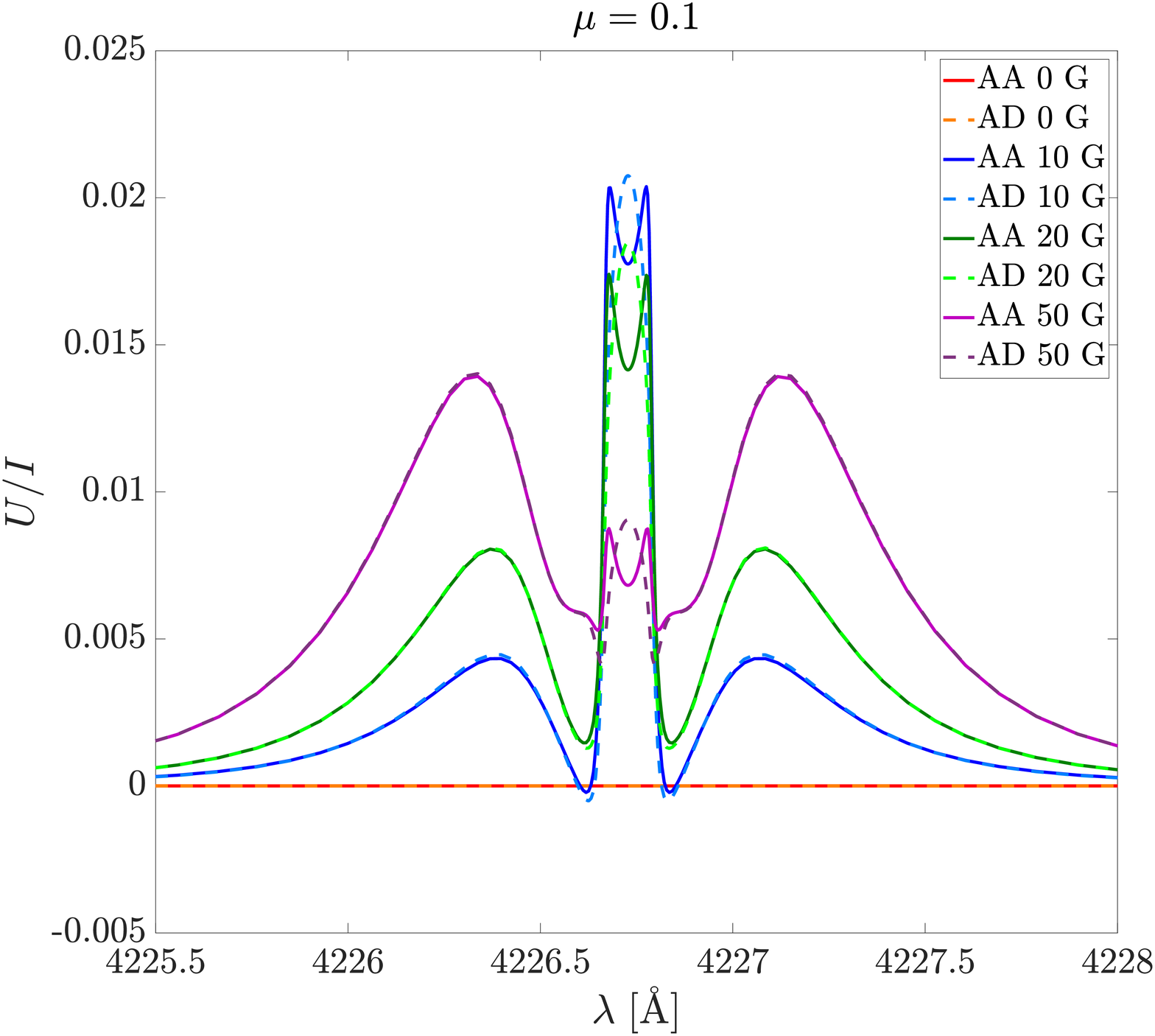}
\caption{Emergent $Q/I$ (left panel) and $U/I$ (right panel) 
calculated in the FAL-C atmospheric model for a LOS with $\mu = 0.1$, 
taking into account PRD effects both under the AA approximation 
(solid lines) and in the general AD case (dashed lines).
The magnetic field is horizontal and contained in the plane defined by the local vertical and the LOS. 
Different field strengths have been applied: 0\,G (red/orange lines), 10\,G (blue/light blue line), 
20\,G (green/pale green lines), and 50\,G (purple/violet lines).
The reference direction for positive $Q$ is taken parallel to the limb.}
  \label{fig:magnetic_case}
\end{figure*}

Figure~\ref{fig:magnetic_case} shows the impact of the AD treatment on the emergent Stokes 
profiles of the Ca~{\sc i} 4227~{\AA} line, for a LOS with $\mu=0.1$, 
when including magnetic fields.
The direction of the magnetic field is horizontal and contained in the plane defined by the 
local vertical and the LOS, while different field strengths have been considered.
For the considered range of magnetic field strengths
(between 0 and 50\,G), the Hanle effect produces a decrease of the 
amplitude of the line-core peak of the $Q/I$ profile (till the disappearance of the 
peak), while giving rise to a peak in the core of the $U/I$ profile.
A trough is found 
in both peaks when performing AA calculations, while the AD counterparts show a peak.
The same figure allows appreciating the sensitivity of the line wings to the same 
magnetic fields through magneto-optical effects: as the magnetic field increases, the wing lobes of $Q/I$ decrease, 
while similar wing lobes appear in $U/I$. 
At such wing wavelengths, AA and AD calculations show a good agreement,
thus suggesting that the AA approximation could be suitable 
to model and exploit magneto-optical effects 
in the 
scattering polarization wings of the Ca~{\sc i} 4227~{\AA} line.
%
\section{Conclusions}\label{sec:conclusions}
%
This paper presents a quantitative comparison between the AA and AD approaches
in the numerical solution of the NLTE RT problem of polarized radiation,
taking scattering polarization and PRD effects into account.
In particular, the $Q/I$ and $U/I$ Stokes profiles of the 
Ca~{\sc i} 4227~{\AA} spectral line have been modeled and discussed, considering realistic 
1D semi-empirical atmospheric models in the presence of magnetic fields.

The most relevant feature of the polarization signals
obtained when performing AD calculations is the disappearance of the 
artificial trough in the line-core peak of the $Q/I$ and $U/I$ scattering 
polarization profiles obtained when using the AA approximation.
This difference between AA and AD calculations is found for all LOS, 
both in the absence and in the presence of magnetic fields. 
This finding highlights the scientific value of AD calculations, as the corresponding 
line-core peak (which is a key observable for Hanle effect diagnostics) is in much better 
agreement with the observations.
Previous investigations have shown that other strong resonance lines, such as Na~{\sc i} D$_2$, show a trough in the line core peak of the $Q/I$ profile when modeled under the AA approximation \citep[e.g.,][]{belluzzi2015,alsinaballester2021}. It is expected that
also in these cases the trough will disappear when AD calculations are carried out.

Additionally, the numerical results support the validity 
of the AA approximation for modeling
the wings of the $Q/I$ and $U/I$ profiles in the static case.
This validates the investigation on the Ca~{\sc i} 4227~{\AA} line performed by \citet{alsinaballester2018}
under the AA approximation, strengthening the diagnostic potential of magneto-optical effects
for exploring the magnetism of stellar atmospheres.

The numerical approach presented in this paper can be applied to model other 
resonance lines, and can be generalized to a wider range of scenarios, 
such as two-term atomic models, and 3D
atmospheric models with arbitrary magnetic and bulk velocity fields.
In particular, the impact of bulk velocity motions
on the modeling of the scattering polarization 
of resonance lines taking PRD effects into account in the general AD 
case will be subject to an upcoming investigation.
\begin{acknowledgements}
The financial support by the Swiss National Science Foundation (SNSF) through grant CRSII5\_180238 is gratefully acknowledged.
E.A.B. and L.B. gratefully acknowledge financial support by SNSF through grant 200021\_175997.
J.T.B. acknowledges the funding received from the European Research Counsil (ERC)
under the European Union's Horizon 2020 research and innovation programme (ERC Advanced Grant agreement No 742265).
\end{acknowledgements}

\bibliographystyle{aa}
\bibliography{bibfile_cai}

\begin{thebibliography}{63}
\expandafter\ifx\csname natexlab\endcsname\relax\def\natexlab#1{#1}\fi

\bibitem[{{Alsina Ballester} {et~al.}(2016){Alsina Ballester}, {Belluzzi}, \&
  {Trujillo Bueno}}]{alsinaballester2016}
{Alsina Ballester}, E., {Belluzzi}, L., \& {Trujillo Bueno}, J. 2016, \apjl,
  831, L15

\bibitem[{{Alsina Ballester} {et~al.}(2017){Alsina Ballester}, {Belluzzi}, \&
  {Trujillo Bueno}}]{alsinaballester2017}
{Alsina Ballester}, E., {Belluzzi}, L., \& {Trujillo Bueno}, J. 2017, \apj,
  836, 6

\bibitem[{{Alsina Ballester} {et~al.}(2018){Alsina Ballester}, {Belluzzi}, \&
  {Trujillo Bueno}}]{alsinaballester2018}
{Alsina Ballester}, E., {Belluzzi}, L., \& {Trujillo Bueno}, J. 2018, \apj,
  854, 150

\bibitem[{{Alsina Ballester} {et~al.}(2019){Alsina Ballester}, {Belluzzi}, \&
  {Trujillo Bueno}}]{alsinaballester2019}
{Alsina Ballester}, E., {Belluzzi}, L., \& {Trujillo Bueno}, J. 2019, \apj,
  880, 85

\bibitem[{{Alsina Ballester} {et~al.}(2021){Alsina Ballester}, {Belluzzi}, \&
  {Trujillo Bueno}}]{alsinaballester2021}
{Alsina Ballester}, E., {Belluzzi}, L., \& {Trujillo Bueno}, J. 2021, \prl,
  accepted

\bibitem[{{Anusha} {et~al.}(2011){Anusha}, {Nagendra}, {Bianda}, {Stenflo},
  {Holzreuter}, {Sampoorna}, {Frisch}, {Ramelli}, \& {Smitha}}]{anusha2011}
{Anusha}, L.~S., {Nagendra}, K.~N., {Bianda}, M., {et~al.} 2011, \apj, 737, 95

\bibitem[{{Belluzzi} \& {Trujillo Bueno}(2012)}]{belluzzi2012}
{Belluzzi}, L. \& {Trujillo Bueno}, J. 2012, \apjl, 750, L11

\bibitem[{{Belluzzi} \& {Trujillo Bueno}(2014)}]{belluzzi2014}
{Belluzzi}, L. \& {Trujillo Bueno}, J. 2014, \aap, 564, A16

\bibitem[{{Belluzzi} {et~al.}(2015){Belluzzi}, {Trujillo Bueno}, \& {Landi
  Degl'Innocenti}}]{belluzzi2015}
{Belluzzi}, L., {Trujillo Bueno}, J., \& {Landi Degl'Innocenti}, E. 2015, \apj,
  814, 116

\bibitem[{{Belluzzi} {et~al.}(2012){Belluzzi}, {Trujillo Bueno}, \&
  {{\v{S}}t{\v{e}}p{\'a}n}}]{belluzzi2012b}
{Belluzzi}, L., {Trujillo Bueno}, J., \& {{\v{S}}t{\v{e}}p{\'a}n}, J. 2012,
  \apjl, 755, L2

\bibitem[{{Benedusi} {et~al.}(2021){Benedusi}, {Janett}, {Belluzzi}, \&
  {Krause}}]{benedusi2021}
{Benedusi}, P., {Janett}, G., {Belluzzi}, L., \& {Krause}, R. 2021, \aap,
  accepted

\bibitem[{{Bianda} {et~al.}(2011){Bianda}, {Ramelli}, {Anusha}, {Stenflo},
  {Nagendra}, {Holzreuter}, {Sampoorna}, {Frisch}, \& {Smitha}}]{bianda2011}
{Bianda}, M., {Ramelli}, R., {Anusha}, L.~S., {et~al.} 2011, \aap, 530, L13

\bibitem[{{Bianda} {et~al.}(2003){Bianda}, {Stenflo}, {Gandorfer}, \&
  {Gisler}}]{bianda2003}
{Bianda}, M., {Stenflo}, J.~O., {Gandorfer}, A., \& {Gisler}, D. 2003, in
  Astronomical Society of the Pacific Conference Series, Vol. 286, Current
  Theoretical Models and Future High Resolution Solar Observations: Preparing
  for ATST, ed. A.~A. {Pevtsov} \& H.~{Uitenbroek}, 61

\bibitem[{{Bommier}(1997{\natexlab{a}})}]{bommier1997a}
{Bommier}, V. 1997{\natexlab{a}}, \aap, 328, 706

\bibitem[{{Bommier}(1997{\natexlab{b}})}]{bommier1997b}
{Bommier}, V. 1997{\natexlab{b}}, \aap, 328, 726

\bibitem[{{Capozzi} {et~al.}(2020){Capozzi}, {Ballester}, {Belluzzi}, {Bianda},
  {Dhara}, \& {Ramelli}}]{capozzi2020}
{Capozzi}, E., {Ballester}, E.~A., {Belluzzi}, L., {et~al.} 2020, \aap, 641,
  A63

\bibitem[{{Carlin} \& {Bianda}(2017)}]{carlin2017}
{Carlin}, E.~S. \& {Bianda}, M. 2017, \apj, 843, 64

\bibitem[{{Casini} {et~al.}(2017){Casini}, {del Pino Alem{\'a}n}, \& {Manso
  Sainz}}]{casini2017b}
{Casini}, R., {del Pino Alem{\'a}n}, T., \& {Manso Sainz}, R. 2017, \apj, 848,
  99

\bibitem[{{Casini} {et~al.}(2014){Casini}, {Landi Degl'Innocenti}, {Manso
  Sainz}, {Land i Degl'Innocenti}, \& {Landolfi}}]{casini2014}
{Casini}, R., {Landi Degl'Innocenti}, M., {Manso Sainz}, R., {Land i
  Degl'Innocenti}, E., \& {Landolfi}, M. 2014, \apj, 791, 94

\bibitem[{{Collados} {et~al.}(2007){Collados}, {Lagg}, {D{\'\i}az Garc{\'\i}
  A}, {Hern{\'a}ndez Su{\'a}rez}, {L{\'o}pez L{\'o}pez}, {P{\'a}ez
  Ma{\~n}{\'a}}, \& {Solanki}}]{collados2007}
{Collados}, M., {Lagg}, A., {D{\'\i}az Garc{\'\i} A}, J.~J., {et~al.} 2007, in
  ASP Conf. Ser., Vol. 368, The Physics of Chromospheric Plasmas, ed.
  P.~{Heinzel}, I.~{Dorotovi{\v{c}}}, \& R.~J. {Rutten}, 611

\bibitem[{{del Pino Alem{\'a}n} {et~al.}(2020){del Pino Alem{\'a}n}, {Trujillo
  Bueno}, {Casini}, \& {Manso Sainz}}]{delpinoaleman2020}
{del Pino Alem{\'a}n}, T., {Trujillo Bueno}, J., {Casini}, R., \& {Manso
  Sainz}, R. 2020, \apj, 891, 91

\bibitem[{{Dumont} {et~al.}(1977){Dumont}, {Omont}, {Pecker}, \&
  {Rees}}]{dumont1977}
{Dumont}, S., {Omont}, A., {Pecker}, J.~C., \& {Rees}, D. 1977, \aap, 54, 675

\bibitem[{{Faurobert}(1987)}]{faurobert1987}
{Faurobert}, M. 1987, \aap, 178, 269

\bibitem[{{Faurobert}(1988)}]{faurobert1988}
{Faurobert}, M. 1988, \aap, 194, 268

\bibitem[{{Faurobert-Scholl}(1992)}]{faurobert_scholl1992}
{Faurobert-Scholl}, M. 1992, \aap, 258, 521

\bibitem[{{Fontenla} {et~al.}(1993){Fontenla}, {Avrett}, \&
  {Loeser}}]{fontenla1993}
{Fontenla}, J.~M., {Avrett}, E.~H., \& {Loeser}, R. 1993, \apj, 406, 319

\bibitem[{{Gandorfer}(2000)}]{gandorfer2000}
{Gandorfer}, A. 2000, {The Second Solar Spectrum: Volume I} (Hochschulverlag
  AG, ETHZ)

\bibitem[{{Gandorfer}(2002)}]{gandorfer2002}
{Gandorfer}, A. 2002, {The Second Solar Spectrum: Volume II} (Hochschulverlag
  AG, ETHZ)

\bibitem[{Gray(1992)}]{Gray92}
Gray, D.~F. 1992, Observation and Analysis of Stellar Photospheres (Cambridge:
  Cambridge University Press)

\bibitem[{{Hackbusch}(1994)}]{hackbusch1994}
{Hackbusch}, W. 1994, Applied Mathematical Sciences, Vol.~95, Iterative
  Solution of Large Sparse Systems of Equations, 1st edn. (New York: Springer)

\bibitem[{{Ishikawa} {et~al.}(2021){Ishikawa}, {Bueno}, {del Pino Alem{\'a}n},
  {Okamoto}, {McKenzie}, {Auch{\`e}re}, {Kano}, {Song}, {Yoshida}, {Rachmeler},
  {Kobayashi}, {Hara}, {Kubo}, {Narukage}, {Sakao}, {Shimizu}, {Suematsu},
  {Bethge}, {De Pontieu}, {Dalda}, {Vigil}, {Winebarger}, {Ballester},
  {Belluzzi}, {{\v{S}}t{\v{e}}p{\'a}n}, {Ramos}, {Carlsson}, \&
  {Leenaarts}}]{ishikawa2021}
{Ishikawa}, R., {Bueno}, J.~T., {del Pino Alem{\'a}n}, T., {et~al.} 2021,
  Science Advances, 7, eabe8406

\bibitem[{{Janett} {et~al.}(2021){Janett}, {Benedusi}, {Belluzzi}, \&
  {Krause}}]{janett2021a}
{Janett}, G., {Benedusi}, P., {Belluzzi}, L., \& {Krause}, R. 2021, \aap,
  accepted

\bibitem[{{Janett} {et~al.}(2017){Janett}, {Carlin}, {Steiner}, \&
  {Belluzzi}}]{janett2017a}
{Janett}, G., {Carlin}, E.~S., {Steiner}, O., \& {Belluzzi}, L. 2017, \apj,
  840, 107

\bibitem[{{Janett} \& {Paganini}(2018)}]{janett2018a}
{Janett}, G. \& {Paganini}, A. 2018, \apj, 857, 91

\bibitem[{{Janett} {et~al.}(2018){Janett}, {Steiner}, \&
  {Belluzzi}}]{janett2018b}
{Janett}, G., {Steiner}, O., \& {Belluzzi}, L. 2018, \apj, 865, 16

\bibitem[{{Jaume Bestard} {et~al.}(2021){Jaume Bestard}, {Trujillo Bueno},
  {{\v{S}}t{\v{e}}p{\'a}n}, \& {del Pino Alem{\'a}n}}]{jaume_bestard2021}
{Jaume Bestard}, J., {Trujillo Bueno}, J., {{\v{S}}t{\v{e}}p{\'a}n}, J., \&
  {del Pino Alem{\'a}n}, T. 2021, \apj, 909, 183

\bibitem[{{Kano} {et~al.}(2017){Kano}, {Trujillo Bueno}, {Winebarger},
  {Auch{\`e}re}, {Narukage}, {Ishikawa}, {Kobayashi}, {Bando}, {Katsukawa},
  {Kubo}, {Ishikawa}, {Giono}, {Hara}, {Suematsu}, {Shimizu}, {Sakao},
  {Tsuneta}, {Ichimoto}, {Goto}, {Belluzzi}, {{\v{S}}t{\v{e}}p{\'a}n}, {Asensio
  Ramos}, {Manso Sainz}, {Champey}, {Cirtain}, {De Pontieu}, {Casini}, \&
  {Carlsson}}]{kano2017}
{Kano}, R., {Trujillo Bueno}, J., {Winebarger}, A., {et~al.} 2017, \apjl, 839,
  L10

\bibitem[{{Landi Degl'Innocenti} \&
  {Landolfi}(2004)}]{landi_deglinnocenti+landolfi2004}
{Landi Degl'Innocenti}, E. \& {Landolfi}, M. 2004, Astrophysics and Space
  Science Library, Vol. 307, {Polarization in Spectral Lines} (Dordrecht:
  Kluwer Academic Publishers)

\bibitem[{{Leenaarts} {et~al.}(2012){Leenaarts}, {Pereira}, \&
  {Uitenbroek}}]{leenaarts2012}
{Leenaarts}, J., {Pereira}, T., \& {Uitenbroek}, H. 2012, \aap, 543, A109

\bibitem[{Mihalas(1978)}]{mihalas1978}
Mihalas, D. 1978, Stellar Atmospheres, 2nd edn. (San Francisco: W.H.~Freeman
  and Company)

\bibitem[{{Nagendra} {et~al.}(2002){Nagendra}, {Frisch}, \&
  {Faurobert}}]{nagendra2002}
{Nagendra}, K.~N., {Frisch}, H., \& {Faurobert}, M. 2002, \aap, 395, 305

\bibitem[{{Ramelli} {et~al.}(2010){Ramelli}, {Balemi}, {Bianda}, {Defilippis},
  {Gamma}, {Hagenbuch}, {Rogantini}, {Steiner}, \& {Stenflo}}]{ramelli2010}
{Ramelli}, R., {Balemi}, S., {Bianda}, M., {et~al.} 2010, in Society of
  Photo-Optical Instrumentation Engineers (SPIE) Conference Series, Vol. 7735,
  \procspie, 77351Y

\bibitem[{{Rees} \& {Saliba}(1982)}]{rees1982}
{Rees}, D.~E. \& {Saliba}, G.~J. 1982, \aap, 115, 1

\bibitem[{{Sampoorna} \& {Nagendra}(2015)}]{sampoorna2015apj}
{Sampoorna}, M. \& {Nagendra}, K.~N. 2015, \apj, 812, 28

\bibitem[{{Sampoorna} {et~al.}(2011){Sampoorna}, {Nagendra}, \&
  {Frisch}}]{sampoorna2011}
{Sampoorna}, M., {Nagendra}, K.~N., \& {Frisch}, H. 2011, \aap, 527, A89

\bibitem[{{Sampoorna} {et~al.}(2008){Sampoorna}, {Nagendra}, \&
  {Stenflo}}]{sampoorna2008}
{Sampoorna}, M., {Nagendra}, K.~N., \& {Stenflo}, J.~O. 2008, \apj, 679, 889

\bibitem[{{Sampoorna} {et~al.}(2017){Sampoorna}, {Nagendra}, \&
  {Stenflo}}]{sampoorna2017}
{Sampoorna}, M., {Nagendra}, K.~N., \& {Stenflo}, J.~O. 2017, \apj, 844, 97

\bibitem[{{Sampoorna} {et~al.}(2010){Sampoorna}, {Trujillo Bueno}, \& {Landi
  Degl'Innocenti}}]{sampoorna2010}
{Sampoorna}, M., {Trujillo Bueno}, J., \& {Landi Degl'Innocenti}, E. 2010,
  \apj, 722, 1269

\bibitem[{{Seaton}(1962)}]{Seaton62}
{Seaton}, M.~J. 1962, Proceedings of the Physical Society, 79, 1105

\bibitem[{{Stenflo}(1994)}]{stenflo1994}
{Stenflo}, J. 1994, {Solar Magnetic Fields: Polarized Radiation Diagnostics},
  Vol. 189 (Springer)

\bibitem[{{Stenflo}(1982)}]{stenflo1982}
{Stenflo}, J.~O. 1982, \solphys, 80, 209

\bibitem[{{Stenflo} {et~al.}(1980){Stenflo}, {Baur}, \& {Elmore}}]{stenflo1980}
{Stenflo}, J.~O., {Baur}, T.~G., \& {Elmore}, D.~F. 1980, \aap, 84, 60

\bibitem[{{Sukhorukov} \& {Leenaarts}(2017)}]{sukhorukov2017}
{Sukhorukov}, A.~V. \& {Leenaarts}, J. 2017, \aap, 597, A46

\bibitem[{{Supriya} {et~al.}(2012){Supriya}, {Nagendra}, {Sampoorna}, \&
  {Ravindra}}]{supriya2012}
{Supriya}, H.~D., {Nagendra}, K.~N., {Sampoorna}, M., \& {Ravindra}, B. 2012,
  \mnras, 425, 527

\bibitem[{{Supriya} {et~al.}(2013){Supriya}, {Sampoorna}, {Nagendra},
  {Ravindra}, \& {Anusha}}]{supriya2013}
{Supriya}, H.~D., {Sampoorna}, M., {Nagendra}, K.~N., {Ravindra}, B., \&
  {Anusha}, L.~S. 2013, \jqsrt, 119, 67

\bibitem[{{Supriya} {et~al.}(2014){Supriya}, {Smitha}, {Nagendra}, {Stenflo},
  {Bianda}, {Ramelli}, {Ravindra}, \& {Anusha}}]{supriya2014}
{Supriya}, H.~D., {Smitha}, H.~N., {Nagendra}, K.~N., {et~al.} 2014, \apj, 793,
  42

\bibitem[{{Traving}(1960)}]{Traving60}
{Traving}, G. 1960, {\"U}ber die Theorie der Druckverbreiterung von
  Spektrallinien (Karlsruhe: Braun)

\bibitem[{{Trujillo Bueno}(2001)}]{trujillo_bueno2001}
{Trujillo Bueno}, J. 2001, in Astronomical Society of the Pacific Conference
  Series, Vol. 236, Advanced Solar Polarimetry -- Theory, Observation, and
  Instrumentation, ed. M.~{Sigwarth}, 161

\bibitem[{{Trujillo Bueno} {et~al.}(2017){Trujillo Bueno}, {Landi
  Degl'Innocenti}, \& {Belluzzi}}]{trujillo_bueno2017}
{Trujillo Bueno}, J., {Landi Degl'Innocenti}, E., \& {Belluzzi}, L. 2017, \ssr,
  210, 183

\bibitem[{{Trujillo Bueno} {et~al.}(2018){Trujillo Bueno},
  {{\v{S}}t{\v{e}}p{\'a}n}, {Belluzzi}, {Asensio Ramos}, {Manso Sainz}, {del
  Pino Alem{\'a}n}, {Casini}, {Ishikawa}, {Kano}, {Winebarger}, {Auch{\`e}re},
  {Narukage}, {Kobayashi}, {Bando}, {Katsukawa}, {Kubo}, {Ishikawa}, {Giono},
  {Hara}, {Suematsu}, {Shimizu}, {Sakao}, {Tsuneta}, {Ichimoto}, {Cirtain},
  {Champey}, {De Pontieu}, \& {Carlsson}}]{trujillo_bueno2018}
{Trujillo Bueno}, J., {{\v{S}}t{\v{e}}p{\'a}n}, J., {Belluzzi}, L., {et~al.}
  2018, \apjl, 866, L15

\bibitem[{{Uitenbroek}(2001)}]{uitenbroek2001}
{Uitenbroek}, H. 2001, \apj, 557, 389

\bibitem[{{Uns\"old}(1955)}]{Unsold55}
{Uns\"old}, A. 1955, {Physik der Sternatmospharen, mit besonderer
  Ber\"ucksichtigung der Sonne.} ({Berlin}: {Springer})

\bibitem[{\v{S}t\v{e}p\'an(2015)}]{stepan2015}
\v{S}t\v{e}p\'an, J. 2015, in IAU Symposium, Vol. 305, Polarimetry, ed. K.~N.
  {Nagendra}, S.~{Bagnulo}, R.~{Centeno}, \& M.~{Jes{\'u}s Mart{\'{\i}}nez
  Gonz{\'a}lez}, 360--367

\end{thebibliography}

\end{document}